\documentstyle[preprint,revtex]{aps}
\begin{document}

\unitlength 0.6cm
\thicklines

\begin{flushright}CEBAF-TH-93-18\end{flushright}
\begin{title}
Quasi-Two-Body Decays of Nonstrange Baryons
\end{title}
\author{Simon Capstick}
\begin{instit}
\centerline{Continuous Electron Beam Accelerator Facility} \\
\centerline{12000 Jefferson Avenue, Newport News, VA 23606, USA.}
\end{instit}
\author{Winston Roberts}
\begin{instit}
Department of Physics, Old Dominion University, Norfolk, VA 23529, USA \\
\centerline{and} \\
\centerline{Continuous Electron Beam Accelerator Facility} \\
\centerline{12000 Jefferson Avenue, Newport News, VA 23606, USA.}
\end{instit}

\begin{abstract}
We examine the decays of nonstrange baryons to the final states
$\Delta\pi$, $N\rho$, $N\eta$, $N\eta^\prime$, $N\omega$,
$N{\textstyle{1\over2}}^+(1440)\pi$,
and
$\Delta{\textstyle{3\over2}}^+(1600)\pi$, in a relativized pair-creation
($^3P_0$) model which
has been
developed in a previous study of the $N\pi$ decays of the same baryon states.
As it
is our goal to provide a guide for the possible discovery of new baryon states
at CEBAF and elsewhere, we examine the decays of resonances which have already
been seen in the partial-wave analyses, along with those of states which are
predicted by the
quark model but which remain undiscovered. The level of agreement between
our calculation and the available widths from the partial-wave analyses is
encouraging.
\end{abstract}
\pacs{13.30.Eg,12.40Qq,14.20Gk}
\newpage
\def\slash#1{#1 \hskip -0.5em / }

\section{Introduction}

The $N^*$ program at the Continuous Electron Beam Accelerator Facility (CEBAF)
is designed to study many interesting aspects of the spectrum and decays of
baryon states.
It contains, among others, experiments to search for the
so-called missing non-strange baryons, to examine the structure of the Roper
resonance [$N^*(1440)$], the
$\Lambda(1405)$ and other low-lying resonances, to search for exotic states,
and to study the couplings of nucleons to strange matter.

Most of these experiments are planned for CEBAF's Hall B, where the CEBAF Large
Acceptance Spectrometer
(CLAS) will be located. The design goals of the CLAS
include good momentum resolution (typically $\delta p/p\approx 1\%$), which is
necessary for
missing mass measurement of particles
like $\pi^0$, $\eta$ and $\omega$. This detector also allows good particle
identification so
that electrons, pions, kaons,
protons and deuterons can be separated, as well as neutrals such as photons and
neutrons.

Some of these experiments, in particular \cite{93033}, will examine the process
$\gamma p \to p\pi^+\pi^-$, with the analysis
focusing on $\gamma p\to \Delta^{++}\pi^-$, $\gamma p\to \Delta^{0}\pi^+$, and
$\gamma p\to p\rho^0$. This experiment proposes to
search for new (missing and undiscovered~\cite{miss-undisc}) resonances with
masses between
the $\Delta\pi$ threshold (1.3 GeV) and 2.3 GeV.
There are other experiments that propose to search in other channels. These
include one~\cite{89039} which will focus on the decays of such states to
$N\eta$, one~\cite{91008}
which will examine
the $N\eta$ and $N\eta^\prime$ channels, and one~\cite{91024} which proposes to
measure $N\omega$ decays.
The three channels $N\eta$,
$N\eta^\prime$ and $N\omega$ offer the advantage of being isospin selective, in
that only
$I={\textstyle{1\over2}}$ $N^*$  resonances (as opposed to
$I={\textstyle{3\over2}}$ $\Delta^*$ resonances) can couple to these final
states. Channels
with the $\Delta(1232)$
substituted for the final-state nucleon may, in
principle, also be investigated, but the increased multiplicity of daughter
hadrons makes
the analysis of such experiments more difficult. However, in \cite{93033}, for
instance, the plan is to
trigger on any event containing one or more charged particles. This means
that multi-particle decays such as $\Delta\eta$ and $\Delta\omega$ will be
present in the data and
can, in principle, be analysed.

A detailed understanding of baryon
physics will also be required for interpretation of the results of many other
CEBAF experiments. Some examples are
experiments to study the
electroproduction of the $\Delta(1232)$ \cite{89037}, the production of baryon
resonances at high momentum transfer \cite{91002},
and experiments which will determine the polarized structure functions of the
nucleon by
electroproduction of $\Delta(1232)$ and $N{\textstyle{1\over2}}^+(1440)$
\cite{91011}. While this list is by no means representative or exhaustive, it
should illustrate the pivotal role that
baryon physics will play in the CEBAF experimental program.

In view of this proposed program, we believe it is important to
focus some attention on models of baryon physics which are capable of providing
detailed information
pertinent to these experiments. In an earlier article~\cite{scwr} we reported
on
a study of the baryon spectrum which
focused on the $N\pi$ couplings of baryon states. Our purpose there was
twofold:
to test an existing model of the
baryon spectrum~\cite{CI} by looking more carefully at its predictions for the
strong decays; and to
investigate a previously proposed solution to the problem of the missing
baryons~\cite{KI}.

Our conclusions were (i) that the model works quite well in describing the
$N\pi$ decays of baryon
resonances, and (ii) that, as suggested in \cite{KI}, the missing baryons
consistently couple weakly to the
$N\pi$ channel, and so should be sought elsewhere. We were also able to
identify aspects of our model
which could be improved.

One possibility for producing the missing baryons is to use photons or
electrons incident
on nucleon targets. The photo- and electroproduction amplitudes
of baryon resonances have been recently examined in this model framework by one
of us~\cite{Cpc}.
Complementary to the work of Refs.~\cite{scwr} and \cite{Cpc} is an examination
of the strong
couplings of the non-strange baryons to decay channels other than $N\pi$. The
motivation for such a study
should be clear from the above description of some aspects of
the CEBAF $N^*$ program; baryon resonances produced in the photo- or
electroexcitation of a nucleon must decay,
and while some of these states will decay to $N\pi$, other channels must be
explored for the missing resonances.
In particular, channels with more than a single
daughter pion may offer some of the best opportunities for their discovery.

The multi-pion final states we study are those in which the extra pion(s)
result from decay of a baryon
resonance to the so-called quasi-two-body final states $N\rho$, $\Delta\pi$,
$N{\textstyle{1\over2}}^+(1440)\pi$ and
$\Delta {\textstyle{3\over2}}^+(1600)\pi$.
In addition, we examine the $N\eta$, $N\eta^\prime$ and $N\omega$ channels,
which will also receive
experimental attention in the CEBAF program.

The study of baryon spectroscopy and decays has a long history; it is clear
that we cannot refer
to every article in the vast literature on these topics. Instead, we mention a
few
of the more recent papers that deal with the decays which we discuss here. A
more complete
list of references can be found in Refs~\cite{scwr} and \cite{CI}.
Using an elementary pseudoscalar
emission model, Koniuk and Isgur \cite{KI} discuss the $N\pi$, $\Delta\pi$ and
$N\eta$ decay channels, as well as the $\Lambda K$ and
$\Sigma K$ channels. Similar work has been done by Blask
{\it et al.}~\cite{Blask}. Koniuk \cite{Koniuk}
has also examined $N\rho$ and $N\omega$ decays in the approximation of
treating the $\rho$ in the narrow-width limit. In a series of articles,
Stancu and Stassart use a flux-tube-breaking model
to discuss decays to the $N\pi$ \cite{SSNpi}, $N\rho$ \cite{SSNrho}, and
$N\omega$ \cite{SSNomega} channels. We have chosen not to examine the channel
in which
the $\pi\pi$ pair are in a relative $S$-wave, with total
isospin zero; Stassart~\cite{SS1} has modeled such decays by treating the
$\pi\pi$ pair as
resulting from a $\sigma$ pseudo-resonance with a mass of about 600 MeV.

This article is organized as follows. In the next section we briefly recap the
model used to describe the strong decays, and discuss our treatment of
quasi-two-body decays.
In section III we present our strong decay amplitudes for the $\Delta\pi$,
$N\rho$,
$N{\textstyle{1\over2}}^+(1440)\pi$, $\Delta {\textstyle{3\over2}}^+(1600)\pi$,
$N\eta$, $N\eta^\prime$, and
$N\omega$ channels.
To explain our assignment of model states to states seen in analyses of the
experimental
data, we reproduce our model predictions for the
$N\pi$ couplings of all states considered here. Our primary source of partial
widths with which to
compare our model predictions is Manley and Saleski's recent partial-wave
analysis of the $N\pi \pi$
final state~\cite{MANSA}; we have also taken some widths from the Particle
Data Group (PDG)
compilation~\cite{PDG}. In section IV we present our conclusions and outlook.

\section{Decay Amplitudes}

\subsection{The Model}

The models we use to obtain the baryon spectrum and strong decay
amplitudes are described in some detail elsewhere \cite{scwr,CI,RB}.
For completeness we outline these
very briefly here. The baryon spectrum is obtained by solving a
Schr\"odinger-like equation in a Fock space consisting solely of valence
quarks. The Hamiltonian used is
\begin{equation}
H=\sum_i \sqrt{{\bf p}_i^2+m_i^2} + V,
\label{Hspec}
\end{equation}
where $V$ is a relative-position and relative-momentum-dependent potential
which includes the usual confining, Coulomb, hyperfine and spin-orbit terms.
Wave functions are expanded
in a large harmonic oscillator basis, and a diagonalization procedure is used
to find the energy eigenvalues. One consequence of the procedure is that none
of the
terms in the Hamiltonian are treated as perturbations. As discussed in
\cite{scwr}, this has a profound effect on the wave functions obtained.

The spectrum that results is comparable to those obtained using other methods,
including the simpler, non-relativistic approaches, but the wave functions we
use here have
been obtained using a more consistent treatment of all the terms in the
Hamiltonian.
Furthermore, the model is extended, with perhaps surprising
success, to the description of many of the higher lying states.

The strong decays of baryons are treated in a version of the pair-creation
model, specifically the $^3P_0$ model. Our {\it ansatz} for the pair creation
operator is
\begin{eqnarray} \label{amp1}
T&=&-3\gamma\sum_{i,j} \int d {\bf p}_i\/ d {\bf p}_j\/ \delta( {\bf p}_i\/ +
{\bf p}_j\/) C_{ij} F_{ij}\nonumber \\
&\times& \sum_m <1,m;1,-m|0,0> \chi_{ij}^m { \cal Y\/ }_1^{-m}( {\bf p}_i\/
- {\bf p}_j\/ )b_i^{\dagger}( {\bf p}_i\/ ) d_j^{\dagger}( {\bf p}_j\/ ).
\end{eqnarray}
Here, $C_{ij}$ and $F_{ij}$ are the color and flavor wavefunctions of the
created pair, both assumed to be singlet, $\chi_{ij}$ is
the spin-triplet wavefunction of the pair, and ${\cal Y}_1({\bf p}_i - {\bf
p}_j)$ is
the vector harmonic indicating that the pair is in a
relative p-wave. Using the wavefunctions described above, we evaluate the
decay amplitude for the process $A\to B C$ as $M=<BC|T|A>.$ In this form
the model has one parameter, which is the pair creation constant $\gamma$.

\subsection{Quasi-Two-Body Decays}

In our previous treatment of the two-body
decays $A\to BC$, the decay widths were obtained from the amplitudes calculated
in the $^3P_0$ model by using
\begin{equation} \label{FGR}
\Gamma=2\pi\int dk k^2 \left|M(k)\right|^2\delta\left(M_a-E_b(k)-E_c(k)\right),
\end{equation}
where $k$ is the momentum of either daughter hadron in the rest frame of the
parent, and $M(k)$ is the decay amplitude calculated in the $^3P_0$ model.
This eventually leads to
\begin{equation}
\Gamma=2\pi\frac{\left|M(k_0)\right|^2E_b(k_0)E_c(k_0)}{M_a},
\end{equation}
and in our version of the model we make the replacements $E_b\to\tilde M_b$,
$E_c\to\tilde M_c$ and $M_a\to\tilde M_a$,
where
$\tilde M_a$, $\tilde M_b$ and $\tilde M_c$ are the masses of these states in
the weak-binding limit.

We now turn our attention to the quasi-two-body decays of the baryon
resonances,
i.e. the decays of the type $A\to BC$, with the subsequent decay
of one of the daughter hadrons, $B$, say, as illustrated in Figure~\ref{fig1}.
One could
use the prescription described above, which amounts to the narrow width
treatment of both daughter hadrons. Indeed, this has been done by Koniuk and
Isgur~\cite{KI} in their
discussion of $\Delta\pi$ decays, as well as by Koniuk~\cite{Koniuk} in his
treatment of $N\rho$ final states. The treatment of some final states in the
narrow-width limit may, however, paint a somewhat inaccurate picture of
couplings,
particularly for states with masses near the threshold of the decay channel
under consideration.
For instance, the nominal mass of the $N{\textstyle{1\over2}}^+(1710)$
resonance
lies very close to the threshold for decays into $N\rho$, if the $\rho$ is
treated as a narrow resonance, so that the result obtained would be very
dependent
on the mass chosen for the $N{\textstyle{1\over2}}^+(1710)$. Phase space is
very limited and
the
decay width obtained in this way would be very small. Similarly, misleading
results
may be obtained for other states that are close to threshold.

Our approach is to take the width of the daughter hadron into account by
replacing the Dirac $\delta$-function in Eq. (\ref{FGR}). One may regard the
$\delta$-function
as arising from the narrow-width limit of the energy denominator
\begin{equation}
\frac{1}{M_a-E_b-E_c-i\varepsilon}=P\frac{1}{M_a-E_b-E_c}
+i\pi\delta\left(M_a-E_b-E_c\right),
\end{equation}
where $\varepsilon$ is related to the total width of the `unstable' final
state.
For daughter hadrons that are broad, the energy denominator becomes
\begin{equation}
\frac{1}{M_a-E_b-E_c-i\frac{\Gamma_t}{2}}=\frac{M_a-E_b-E_c
+i\frac{\Gamma_t}{2}}{\left(M_a-E_b-E_c\right)^2+\frac{\Gamma_t^2}{4}},
\end{equation}
implying the replacement
\begin{equation}
\delta\left(M_a-E_b-E_c\right)\to\frac{\Gamma_t}
{\pi\left[\left(M_a-E_b-E_c\right)^2+\frac{\Gamma_t^2}{4}\right]}.
\end{equation}
The decay rate for $A\to(X_1X_2)_BC$ then generalizes to~\cite{KokoskiIsgur}
\begin{equation}
\Gamma=\int_0^{k_{\max}} dk\frac{k^2\left|M(k)\right|^2\Gamma_t(k)}
{\left(M_a-E_b(k)-E_c(k)\right)^2+\frac{\Gamma_t(k)^2}{4}},
\end{equation}
where $\Gamma_t(k)$ is the energy-dependent total width of the unstable
daughter hadron, $B$ in this case. Our prescription for this quantity depends
on the
daughter hadron being studied. For states like the $\Delta$ and the $\rho$,
where the hadron decays with a branching fraction of close to 100\% into a
single two-body
final state, we use the energy-dependent width as calculated in our version of
the $^3P_0$ model.
For a state like the Roper resonance $N{\textstyle{1\over2}}^+(1440)$, which
has a branching
fraction of about
70\% to $N\pi$, we write
\begin{equation}
\Gamma_t(k)=\Gamma_{N\pi}(k)+0.3\Gamma_0\left(\frac{k}{k_0}\right)^{2\ell+1}
\frac{k^2+\kappa^2}{k_0^2+\kappa^2},
\label{Gammat}
\end{equation}
where $\Gamma_0$ is the total width of the Roper at the pole position. The
first term is the energy-dependent $N\pi$ width of the Roper, calculated
in the $^3P_0$ model. The second term is the width of the Roper for decays into
other final states such as $N\pi\pi$. Here
$k_0$ is a reference momentum, chosen to
be the momentum of the $\Delta$ in the $\Delta\pi$ decay of the Roper
measured in its rest frame, and $\kappa$ is a phenomenological constant, chosen
to be 0.35 MeV.
The power $2\ell+1$ represents the dependence
of the energy-dependent width on angular momentum, and is chosen here to be
unity. We note that the results we present are largely
insensitive to changes in $\kappa$, $\ell$, and $k_0$, nor are they overly
sensitive to the inclusion of the second term in Eq.~\ref{Gammat}.

The variable of integration in the expression above is $k$, the magnitude of
the three-momentum of the daughter hadron $B$. In the rest frame of $A$, this
ranges
from $k=0$ ($X_1$ and $X_2$ back-to-back, with ${\bf q_1}={\bf -q_2}$) to
$k_{max}$ ($X_1$ and $X_2$ colinear). These limits correspond to
$M_b(k)=M_a-M_c$, and
$M_b(k)=M_{X_1}+M_{X_2}$, respectively, where $M_b(k)$ is the
momentum-dependent effective mass of daughter hadron $B$.

In what follows we present decay amplitudes for decays to $\Delta\pi$, $N\rho$,
$N\eta$, $N\eta^\prime$, $N\omega$, $N{\textstyle{1\over2}}^+(1440)\pi$ and
$\Delta {\textstyle{3\over2}}^+(1600)\pi$.
For most of these channels we use the prescription described above. For
$N\eta$ and $N\omega$,
the unstable daughter hadrons are sufficiently narrow that we
ignore their widths. This is because a total width of less than 10 MeV is well
within the expected
accuracy of the model, so that such states may be safely treated as being
narrow.

We end this subsection with a comparison of the method we use for treating
quasi-two-body decays with other prescriptions found in the literature. In
their
treatment of $N\rho$ decays, Stancu and Stassart~\cite{SSNrho} use the
prescription
of Cutkosky {\it et al.}~\cite{CFHK} by including a relativistic Breit-Wigner
mass distribution
$\sigma$, and integrating over the mass of the $\rho$. Their width to $N\rho$
is then
\begin{equation}
\Gamma=\int_{(2m_\pi)^2}^{(M_R-M_N)^2}dm_\rho^2\sigma(m_\rho^2)
\Gamma_{R\to N\rho}(m_\rho^2),
\end{equation}
with
\begin{equation}
\sigma(m_\rho^2)=\frac{\Gamma_\rho(m_\rho^2)m_\rho/\pi}{(M_\rho^2-m_\rho^2)^2
+\Gamma^2_\rho(m_\rho^2)m_\rho^2},
\end{equation}
and where
\begin{equation}
\Gamma_\rho(m_\rho^2)=\Gamma_0\left(\frac{k}{k_0}\right)^3
\frac{M_\rho}{m_\rho}\frac{2k_0^2}{k^2+k_0^2}
\end{equation}
is the energy dependent width of the $\rho$. This last form is suggested by
Gottfried and Jackson~\cite{Jackson}, with
$M_\rho=770$ MeV, $\Gamma_\rho(M_\rho^2)=153$ MeV, $k_0=k(M_\rho^2)$,
and where $k(m_\rho^2)$ is the relative momentum of the pions in the decay
$\rho\to\pi\pi$. We have compared the
results of using this prescription to our results, and find no significant
differences.

\subsection{The Parameters}

The parameters of the model are the pair-creation strength $\gamma$,
and the gaussian parameters of the meson and baryon wave
functions. For consistency, the parameters that this work has in common with
Ref.~\cite{scwr}
have not been changed from the values used there, so that $\gamma=2.6$,
and the gaussian parameters $\alpha$ of all baryon wave functions are set to
0.5 GeV (a
common value is necessary in a decay calculation in order to maintain
orthogonality of the initial and final baryon wavefunctions). In
addition the corresponding parameters for all the mesons are set to
$\beta$=0.4 GeV.

An additional parameter for this work is $\kappa$, discussed in the previous
subsection.
However, as mentioned there, the results that we present are largely
independent
of $\kappa$; we use a value of 0.35 GeV.

\section{Results}

Our results are presented in the form of several tables. In order to list as
many of our
results as possible in a form that makes various comparisons relatively easy to
carry out,
these tables are necessarily dense; accordingly we have included the following
guide to their
organization.

For each set of nucleon resonances, there are three tables. The first of these
lists the model state,
its decay amplitudes into the $N\pi$, $N\eta$, $N\eta^\prime$ channels, and its
helicity partial-wave
 and total decay amplitudes for the $N\omega$ channel. The second table lists
the helicity partial-wave and total decay amplitudes for each state in the
$\Delta\pi$ and
$N\rho$ channels, while for the nucleons beyond the $N=2$ band, the third table
lists the corresponding quantities for the $\Delta(1600)\pi$ and $N(1440)\pi$
channels (we do not
present the latter type of table for nucleons in the $N\le 2$ bands). The
format for presentation
of the decay amplitudes for the $\Delta$ resonances is similar, although we
have split
some of the larger tables. In addition to this division among the tables, for
each state
there are two rows of entries. The first row lists our model
predictions, the second the values published in the recent $N\pi\pi$
partial-wave analysis of
Manley and Saleski \cite{MANSA}, along with the Particle Data Group~\cite{PDG}
name, $N\pi$ partial wave,
star rating, and $N\pi$ amplitude for this state.
In all cases, the predicted $N\pi$ amplitudes in the first row are the same as
in Ref.~\cite{scwr},
reproduced here for ease of comparison. The association of a given state from
the partial-wave analyses
with a model state in our tables is designed to make explicit our assigment of
model states
to established states.

For all but the $N\pi$ channels, we give the sign of the amplitude along with
its
magnitude. This sign must be understood as the sign relative to the
unmeasurable sign of the
$N\pi$ production amplitude. In other words, the signs presented with our model
amplitudes
are the product $\zeta_{N\pi}\zeta_{BM}$ of the sign for the $N\pi$ entrance
channel times that of the decay channel
being considered (one can extract the sign for the amplitude with the
$\gamma N$ entrance
channel by taking the product of the sign $\zeta_{\gamma N}\zeta_{N\pi}$ for
the photocoupling amplitudes from
Ref.~\cite{Cpc} and the signs in the tables in this paper).

All theoretical amplitudes are also given with upper and lower limits, along
with the central value,
in order to convey the uncertainty in our results due to the uncertainty in the
resonance's mass. These
correspond to our predictions for the amplitudes for a resonance whose mass is
set
to the upper and lower limits, and to the central value, of the experimentally
determined mass.
For states as yet unseen in the analyses of the data, we have adopted a
`standard' uncertainty in the mass of
150 MeV and used the model predictions for the state's mass as the central
value.

By examining the tables one can see the significant mass dependence
in some of the decay amplitudes. One example
is the $N\rho$ decay of the $N(1520)D_{13}$ (Table II). At its central
mass value, the amplitude for this decay is 2.5 MeV$^{{\textstyle{1\over2}}}$,
while if the mass is increased by the uncertainty from the partial-wave
analyses
(150 MeV
in this case), the amplitude becomes 9.0 MeV$^{{\textstyle{1\over2}}}$,
corresponding to a factor of more than 12 in the decay width. Although
this is an extreme example, many amplitudes exhibit similar
dependencies.

\subsection{$N^*$ Resonances in the $N\le 2$ Bands}

The model predictions for nucleon resonances in the $N\le 2$ bands
(Tables \ref{nNle2PS} and \ref{nNle2Npipi})
are in generally good agreement with the analyses of the
experimental data. The larger predicted amplitudes correspond to
larger measured amplitudes, although the magnitudes of the theoretical and
measured amplitudes
may differ. Predicted signs of amplitudes are also mostly in agreement with the
experimentally
reported signs; most of those that are in disagreement with their
experimental counterparts correspond to small measured amplitudes. Notable
exceptions are
the $s$- and $d$-wave $\Delta\pi$ partial amplitudes of $D_{13}(1700)$
(although we successfully predict a relatively
large total width to $\Delta\pi$ for this state), and the $N\rho$ amplitudes
of the $P_{13}(1720)$ and
$F_{15}(2000)$ resonances. In addition, Manley and Saleski~\cite{MANSA} also
find a second light $P_{13}$ resonance,
$P_{13}(1880)$, in the $N\rho$ channel. In our model, none of the missing
$P_{13}$ states in the $N=2$ band (whose masses
are in this region) are predicted to have large $N\rho$ widths.

As far as the missing states are concerned, we predict sizeable amplitudes in
the
$\Delta\pi$ channel for the predicted states
$[N{\textstyle{1\over2}}^+]_4(1880)$, $[N{\textstyle{1\over2}}^+]_5(1975)$,
$[N{\textstyle{3\over2}}^+]_3(1910)$,
$[N{\textstyle{3\over2}}^+]_4(1950)$,
$[N{\textstyle{3\over2}}^+]_5(2030)$, and $[N{\textstyle{5\over2}}^+]_3(1995)$.
Thus,
our model predicts that these states should be clearly seen in the experiment
proposed in Ref.~\cite{93033},
and can be viewed as a guide to which channels and partial waves show the
greatest
potential for their discovery. Note in Tables~\ref{nNle2PS} and
\ref{nNle2Npipi}
we have also reassigned the
two-star state $N(2000)F_{15}$ from
the model state $[N{\textstyle{5\over2}}^+]_2(1980)$ to
$[N{\textstyle{5\over2}}^+]_3(1995)$, on the basis of
its $\Delta\pi$ and $N\rho$ decays; the
assignment from Ref.~\cite{scwr} based on the (small) $N\pi$ amplitudes of
these
model states was at best
tentative.

All of the $\Delta(1600)\pi$ and $N(1440)\pi$ amplitudes for these states are
small,
the largest corresponding to a width of about 16 MeV. Thus we do not show
tables for these
channels in this sector. This result implies little hope of discovering missing
baryon resonances
in either of these two channels. We note, however, that the small amplitudes
in
the $N(1440)\pi$ channel are in contradiction with some large amplitudes
reported in
Ref.~\cite{MANSA}, and may point to possible shortcomings in the model. In
particular,
all of our predictions, but especially the couplings to broad final states like
Roper $\pi$, may be modified by the inclusion of
decay-channel-coupling
effects in the spectrum and wavefunctions, which we plan to address in a
later study.

In the $N\eta$, $N\eta^\prime$ and $N\omega$ channels, there are not many
extracted
amplitudes published.
Our prediction for the $S_{11}(1535)$ is compatible with the measured
amplitude, but that
for its heavier counterpart $S_{11}(1650)$ is too large. Note, however, that
there is
recent controversy about existing analyses of the $N\eta$
final-state~\cite{BNefkens}. Most states in this
sector have little or no phase space for decays to these three channels, so it
is not
surprising to find small amplitudes. The exceptions to this are the states
$[N{\textstyle{3\over2}}^+]_2(1870)$,
$[N{\textstyle{3\over2}}^+]_3(1910)$, and
$[N{\textstyle{3\over2}}^+]_4(1950)$, all of which couple strongly
to $N\omega$, and slightly less
strongly to $N\eta$. This suggests that these channels offer good opportunities
for the
discovery or confirmation of these states; they should therefore be seen in the
experiments
proposed in Refs.~\cite{89039,91008,91024}.

\subsection{$\Delta$ Resonances in the $N\le 2$ Bands}

The magnitudes and signs of most $\Delta\pi$ and $N\rho$ amplitudes are well
reproduced
in this sector (see Tables~\ref{dNle2Npipi1} and~\ref{dNle2Npipi2}). The most
notable
exceptions are
the signs of the $s_{{\textstyle{1\over2}}}$ $N\rho$ partial
amplitude of the $S_{31}(1620)$, and of the $f_{{\textstyle{3\over2}}}$ $N\rho$
partial
amplitude of
$F_{37}(1950)$. Our calculation suggests that the new state
$\Delta(1740)P_{31}$ found
by Manley and Saleski~\cite{MANSA} is the formerly missing first
$\Delta{\textstyle{1\over2}}^+$ state; we predict a sizeable
$N\rho$ amplitude for this state in the partial wave in which it was
discovered. However,
we also predict a sizeable $\Delta\pi$ amplitude for this state, whereas
Manley and Saleski's
analysis shows no evidence for such a state in this channel. Our results also
suggest that the other
missing state here, the fourth $\Delta{\textstyle{3\over2}}^+$, should be
visible in both the
$\Delta\pi$ and
$N\rho$ channels.

This sector also contains a puzzle, which is the $F_{35}(1750)$ state reported
in Ref.~\cite{MANSA}.
The lowest lying model state in this sector is the
$[\Delta{\textstyle{5\over2}}^+]_1$ state
at 1910 MeV.
On the basis of our calculated $N\pi$, $\Delta\pi$, and $N\rho$ amplitudes,
this state corresponds most closely to $F_{35}(1905)$.
The three-quark spectrum therefore can not easily accomodate the state
$F_{35}(1750)$, unless
all of the $J^P={\textstyle{5\over2}}^+$ $\Delta$'s are reassigned. This is
also problematic
since there are just
two states in the $N=2$ band in the nonrelativistic quark model, and one of
these three states would have to be reassigned
to the $N=4$ band, for which it is significantly too light.

As is the case with the nucleons in these bands, the couplings of the
$\Delta$'s to the
$\Delta(1600)\pi$ and $N(1440)\pi$ channels are all small, with the largest
partial width
into one of these channels being around 16 MeV.

\subsection{$N^*$ Resonances in the $N=3$ Band}

The number of widths extracted from the data with which we can compare our
model diminishes
significantly as we increase the masses of the baryon resonances we consider,
and this begins
to be apparent in this band. There are few extracted widths for these nucleons
in the $\Delta\pi$ and
$N\rho$ channels (Table~\ref{nNeq3Npipi})
and none in the $N\eta$, $N\eta^\prime$ and $N\omega$ channels
(Table~\ref{nNeq3PS}). In the $\Delta\pi$ and
$N\rho$ channels, the agreement of the model with the partial-wave analysis of
Manley and Saleski
is comparable to that obtained in
other sectors. Discrepancies occur in the description of the $N\rho$ partial
amplitudes
of $S_{11}(2090)$, and in the signs of the amplitudes of the $\Delta\pi$ and
$N\rho$
amplitudes of the $D_{13}(2080)$, although in both cases the total widths are
in reasonable agreement
with experiment.

Many of the as-yet-unseen states couple strongly to the $\Delta\pi$ and $N\rho$
channels (Table~\ref{nNeq3Npipi}).
The most noticeable of these is the predicted state
$[N{\textstyle{1\over2}}^-]_5(2070)$ which
has a $\Delta\pi$
amplitude in excess of 13 MeV$^{{\textstyle{1\over2}}}$. Thus, these channels
offer good
opportunities for
discovery of many of these missing resonances.

In the $N\eta$, $N\eta^\prime$ and $N\omega$ channels (Table~\ref{nNeq3PS}),
some undiscovered states have appreciable
couplings, but few of these will be likely to yield what may be termed
`smoking-gun' signals, with the
possible exceptions of the states $[N{\textstyle{1\over2}}^-]_5(2070)$ and
$[N{\textstyle{7\over2}}^-]_2(2205)$ in the $N\omega$ channel.
Our results (Table~\ref{nNeq3Npipi2}) indicate that weak evidence
for the existence of the lowest-lying $N=3$ band resonances
$N{\textstyle{1\over2}}^-(2090)S_{11}$, $N{\textstyle{3\over2}}^-(2080)D_{13}$,
and $N{\textstyle{3\over2}}^-(2200)D_{15}$ could be strengthened with
$\Delta(1600)\pi$ or
$N(1440)\pi$ experiments
in this mass range. In the $\Delta(1600)\pi$ channel, a few undiscovered states
have appreciable couplings, most
noticeably
$[N{\textstyle{3\over2}}^-]_5(2095)$, $[N{\textstyle{5\over2}}^-]_2(2080)$ and
$[N{\textstyle{3\over2}}^-]_4(2055)$. In the
$N(1440)\pi$ channel, only the model states
$[N{\textstyle{1\over2}}^-]_5(2070)$ and $[N{\textstyle{5\over2}}]_3^-(2095)$
have appreciable couplings.

\subsection{$\Delta$ Resonances in the $N\ge 3$ Bands}

In this sector, there is general agreement between our model and the
partial-wave analyses
in the $\Delta\pi$ channel (Table \ref{dNge3NDpi}), although there are few
extracted
widths with which to compare. However, the $N\rho$
couplings (Table \ref{dNge3Nrho}) of $S_{31}(1900)$ are not well reproduced.
Most of the states
in this sector may be classified as undiscovered, and most of these have
sizeable partial
widths in the $\Delta\pi$ and $N\rho$ channels. These channels
should, therefore, allow for the discovery of many new baryon states.
In the $\Delta(1600)\pi$ channel (Table~\ref{dNge2Npipi2}), only the
$[\Delta{\textstyle{3\over2}}^-]_3(2145)$ offers a
good possibility for discovery, although it should be possible to confirm the
existence of the
state $\Delta{\textstyle{1\over2}}^-(1900)S_{31}$, and the weakly-established
states
$\Delta{\textstyle{3\over2}}^-(1940)D_{33}$
and $\Delta{\textstyle{7\over2}}^+(2390)F_{37}$ which appear in the $N\pi$
analyses~\cite{PDG}. In the $N(1440)\pi$ channel,
we predict no real possibilities for discovering new baryons, although
confirmation of the
tentatively established states $\Delta{\textstyle{1\over2}}^-(2150)S_{31}$ and
$\Delta{\textstyle{7\over2}}^+(2390)F_{37}$ should
be possible.

\subsection{$N^*$ resonances in the $N\ge 4$ Bands}

Very few of this group of states predicted by the quark model have been seen;
our results (Table~\ref{nNge4Npipi})
indicate that many of these states should be discovered in the $\Delta\pi$ and
$N\rho$ channels.
Partial widths into the $N\eta$, $N\eta^\prime$ and $N\omega$ channels
(Table~\ref{nNge4PS}) are generally small.
The two lightest $N=4$ band $N{\textstyle{1\over2}}^+$ states both have
sizeable couplings to
the $\Delta(1600)\pi$
channel (Table~\ref{nNge4Npipi2}); we have assigned the lightest of these to
the tentative state
$N{\textstyle{1\over2}}^+(2100)P_{11}$ from the $N\pi$ analyses. Verification
of the existence of
this state, and discovery of the second
state ($[N{\textstyle{1\over2}}^+]_7(2210)$ in our model) appear to be possible
in this
channel. None of the states we
have considered in this sector, with the exception of the
$N{\textstyle{1\over2}}^+(2100)P_{11}$,
have appreciable couplings to
$N(1440)\pi$.

\section{Conclusions and Outlook}

Our results indicate that many of the baryon states predicted by the quark
model but as yet
unseen in the partial-wave analyses should appear first in analyses of decays
to the $N\pi\pi$ channel
through the $\Delta\pi$ and $N\rho$ quasi-two-body channels. The rough
agreement of the
signs and magnitudes of our predicted amplitudes with the majority of the
existing data
in these channels also suggests that our predictions can act as a rough guide
to the specific channel,
partial wave, and mass range in which to look for these new states. The same is
also true for
the lighter missing and undiscovered $N$ ($I={\textstyle{1\over2}}$) states in
the $N\eta$,
$N\eta^\prime$, and $N\omega$
channels. These amplitudes tend to be quite small for the higher-mass states
studied here.

As mentioned above, our model tends to predict small amplitudes for the lighter
($N\leq 2$) baryons to
decay to $N(1440)\pi$
and $\Delta(1600)\pi$. This can be traced back to the node in the radial part
of the momentum-space
wavefunctions of these final baryon states, which are described as radial
excitations in our model. We
expect that this aspect of our model, which disagrees with some of Manley and
Saleski's results~\cite{MANSA},
will be affected by including decay-channel couplings into the spectrum and
wavefunctions (note that $N(1440)$ is an exceptionally broad state
in our model and in the recent analyses). We intend to address this problem in
a later
study. We have, however, displayed some results from our calculation for decays
of heavier baryons into these final
states, as in this case there can be appreciable amplitudes; it is possible
that the larger amplitudes
will remain large after the necessary corrections are applied.

\figure{The quasi-two-body decay $A\to (X_1X_2)_BC$, showing
the kinematic variables.\label{fig1}}
\squeezetable
\begin{table}
\caption{Results for nucleons in the $N=1$ and $N=2$ bands in the $N\pi$,
$N\eta$, $N\eta^\prime$, and
$N\omega$ channels. Notation for model states is $[J^P]_n({\rm mass[MeV]})$,
where $J^P$ is the
spin/parity of the state and $n$ its principal quantum number. The first row
gives our model results,
while the second row lists the corresponding numbers obtained by Manley and
Saleski in their partial-wave analysis,
as well as the
Particle Data Group name, $N\pi$ partial wave, star rating, and $N\pi$
amplitude for the state.}
\label{nNle2PS}
\begin{tabular}{lrrrrrrr}
\multicolumn{1}{l} {model state}
& \multicolumn{1}{r}{$N\pi$}
& \multicolumn{1}{r}{$N\eta$}
& \multicolumn{1}{r}{$N\eta^\prime$}
& \multicolumn{1}{r}{$N\omega$}
& \multicolumn{1}{r}{$N\omega$}
& \multicolumn{1}{r}{$N\omega$}
& \multicolumn{1}{r}{$\sqrt{\Gamma^{\rm tot}_{N\omega}}$}\\
\multicolumn{1}{l} {$N\pi$ state/rating} & & & & & & &\\
\tableline
$[N{\textstyle{1\over2}}^-]_1(1460)$ & $14.7\pm 0.5$ & $+14.6^{+0.7}_{-1.3}$ &
& & & &\\
$N(1535)S_{11}$**** & $8.0\pm 2.8$ & $+8.1\pm 0.8$ & & & &\\

$[N{\textstyle{1\over2}}^-]_2(1535)$ & $12.2\pm 0.8$ & $-7.8^{+0.1}_{-0.0}$ & &
& & &\\
$N(1650)S_{11}$**** & $8.7\pm 1.9$ & $-2.4\pm 1.6$ & & & &\\

& & & & $d_{\textstyle{1\over2}}$ & $s_{\textstyle{3\over2}}$ &
$d_{\textstyle{3\over2}}$ &\\
\cline{5-7}
$[N{\textstyle{3\over2}}^-]_1(1495)$ & $8.6\pm 0.3$ & $+0.4^{+2.9}_{-0.4}$ & &
& & &\\
\end{tabular}
\end{table}
\begin{table}
\caption{Results for nucleons in the $N=1$ and $N=2$ bands in the $\Delta\pi$,
and $N\rho$ channels.
Notation for model states is $[J^P]_n({\rm mass[MeV]})$, where $J^P$ is the
spin/parity of the state
and $n$ its principal quantum number. The first row gives our model results,
while the second row
lists the numbers obtained by Manley and Saleski in their analysis, as well as
the
Particle Data Group name, $N\pi$ partial wave, and star rating for the state.}
\label{nNle2Npipi}
\begin{tabular}{lrrrrrrr}
\multicolumn{1}{l} {model state}
& \multicolumn{1}{r}{$\Delta\pi$}
& \multicolumn{1}{r}{$\Delta\pi$}
& \multicolumn{1}{r}{$\sqrt{\Gamma^{\rm tot}_{\Delta\pi}}$}
& \multicolumn{1}{r}{$N\rho$}
& \multicolumn{1}{r}{$N\rho$}
& \multicolumn{1}{r}{$N\rho$}
& \multicolumn{1}{r}{$\sqrt{\Gamma^{\rm tot}_{N\rho}}$}\\
\multicolumn{1}{l} {$N\pi$ state/rating} & & & & & & &\\
\tableline
& $d$ & & & $s_{\textstyle{1\over2}}$ & $d_{\textstyle{3\over2}}$ & &\\
\cline{2-2}
\cline{5-6}
$[N{\textstyle{1\over2}}^-]_1(1460)$ & $+1.4\pm 0.3$ & $$ & $1.4\pm 0.3$ &
$-0.7\pm 0.1$
& $+0.4\pm 0.1$ & $$  & $0.8^{+0.2}_{-0.1}$\\
$N(1535)S_{11}$**** & $0.0$ & & $0.0$ & $-1.7\pm 0.6$ & $-1.4\pm 0.7$ &
& $2.2\pm 0.6$\\

$[N{\textstyle{1\over2}}^-]_2(1535)$ & $+3.6^{+0.8}_{-0.6}$ & $$ &
$3.6^{+0.8}_{-0.6}$
& $+0.9^{+0.3}_{-0.2}$ & $+0.4\pm 0.1$ & $$ & $1.0^{+0.3}_{-0.2}$ \\
$N(1650)S_{11}$**** & $+1.7\pm 0.6$ & & $1.7\pm 0.6$ & $0.0$ & $+2.2\pm 0.9$ &
& $2.2\pm 0.9$\\

& $s$ & $d$ & & $d_{\textstyle{1\over2}}$ & $s_{\textstyle{3\over2}}$ &
$d_{\textstyle{3\over2}}$ &\\
\end{tabular}
\end{table}
\vbox{\begin{table}
\caption{Results for $\Delta$ states in the $N=0$, $N=1$, and $N=2$ bands in
the $N\pi$ and $\Delta\pi$ channels. Notation as in Table~\ref{nNle2PS}.}
\label{dNle2Npipi1}
\begin{tabular}{lrrrr}
\multicolumn{1}{l} {model state}
& \multicolumn{1}{r}{$N\pi$}
& \multicolumn{1}{r}{$\Delta\pi$}
& \multicolumn{1}{r}{$\Delta\pi$}
& \multicolumn{1}{r}{$\sqrt{\Gamma^{\rm tot}_{\Delta\pi}}$}\\
\multicolumn{1}{l} {$N\pi$ state/rating} & & & &\\
\tableline
& & $p$ & &\\
\cline{3-3}
$[\Delta{\textstyle{1\over2}}^-]_1(1555)$ & $5.1\pm 0.7$ & $-4.2^{+1.3}_{-1.8}$
& $$
& $4.2^{+1.8}_{-1.3}$\\
$\Delta(1620)S_{31}$**** & $6.5\pm 1.0$ & $-9.7\pm 1.3$ & $$ & $9.7\pm 1.3$\\

& & $s$ & $d$ &\\
\cline{3-4}
$[\Delta{\textstyle{3\over2}}^-]_1(1620)$ & $4.9\pm 0.7$ &
$+15.4^{+0.9}_{-1.0}$
& $+5.0^{+2.4}_{-1.8}$ & $16.2^{+1.7}_{-1.5}$\\
$\Delta(1700)D_{33}$**** & $6.5\pm 2.0$ & $+21.1\pm 4.7$ & $+5.1\pm 2.2$
& $21.7\pm 4.6$\\
\end{tabular}
\end{table}}
\vbox{\begin{table}
\caption{Results for $\Delta$ states in the $N=0$, $N=1$, and $N=2$ bands in
the $N\rho$ channel. Notation as in Table~\ref{nNle2Npipi}.}
\label{dNle2Npipi2}
\begin{tabular}{lrrrr}
\multicolumn{1}{l} {model state}
& \multicolumn{1}{r}{$N\rho$}
& \multicolumn{1}{r}{$N\rho$}
& \multicolumn{1}{r}{$N\rho$}
& \multicolumn{1}{r}{$\sqrt{\Gamma^{\rm tot}_{N\rho}}$}\\
\multicolumn{1}{l} {$N\pi$ state/rating} & & & &\\
\tableline
& $s_{\textstyle{1\over2}}$ & $d_{\textstyle{3\over2}}$ & &\\
\cline{2-3}
$[\Delta{\textstyle{1\over2}}^-]_1(1555)$ & $-3.6^{+1.3}_{-2.5}$ &
$-0.3^{+0.1}_{-0.2}$ & $$
& $3.6^{+2.5}_{-1.3}$\\
$\Delta(1620)S_{31}$**** & $+6.2\pm 0.9$ & $-2.4\pm 0.2$ & $$ & $6.6\pm 0.8$\\

& $d_{\textstyle{1\over2}}$ & $s_{\textstyle{3\over2}}$ &
$d_{\textstyle{3\over2}}$ &\\
\cline{2-4}
$[\Delta{\textstyle{3\over2}}^-]_1(1620)$ & $-1.2^{+0.6}_{-1.2}$
& $+3.4^{+2.2}_{-1.7}$ & $+0.5^{+0.5}_{-0.2}$ & $3.6^{+2.5}_{-1.8}$\\
$\Delta(1700)D_{33}$**** & $$ & $+6.8\pm 2.3$ & $$ & $6.8\pm 2.3$\\
\end{tabular}
\end{table}}
\begin{table}
\caption{Results for the lightest few negative-parity nucleon resonances of
each $J$ in
the N=3 band in the $N\pi$, $N\eta$, and $N\omega$ channels. Notation as in
Table~\ref{nNle2PS}.}\label{nNeq3PS}
\begin{tabular}{lrrrrrrr}
\multicolumn{1}{l} {model state}
& \multicolumn{1}{r}{$N\pi$}
& \multicolumn{1}{r}{$N\eta$}
& \multicolumn{1}{r}{$N\eta^\prime$}
& \multicolumn{1}{r}{$N\omega$}
& \multicolumn{1}{r}{$N\omega$}
& \multicolumn{1}{r}{$N\omega$}
& \multicolumn{1}{r}{$\sqrt{\Gamma^{\rm tot}_{N\omega}}$}\\
\multicolumn{1}{l} {$N\pi$ state/rating}\\
\tableline
& & & & $s_{\textstyle{1\over2}}$ & $d_{\textstyle{3\over2}}$ &\\
\cline{5-6}
$[N{\textstyle{1\over2}}^-]_3(1945)$ & $5.7^{+0.5}_{-1.6}$ &
$+2.4^{+1.5}_{-2.3}$
& $-3.6^{+3.3}_{-2.3}$ & $-0.9^{+3.0}_{-2.2}$ & $-5.6^{+1.8}_{-0.7}$ &
& $5.7\pm 1.3$\\
$N{\textstyle{1\over2}}^-(2090)S_{11}$* & $7.9\pm 3.8$\\
& \\

$[N{\textstyle{1\over2}}^-]_4(2030)$ & $3.7^{+0.5}_{-1.1}$ &
$-1.0^{+1.5}_{-1.1}$
& $+1.3\pm 1.3$ & $-0.1^{+2.4}_{-1.9}$ & $-2.8^{+1.2}_{-0.6}$ &
& $2.8^{+1.1}_{-0.0}$\\

$[N{\textstyle{1\over2}}^-]_5(2070)$ & $2.1^{+0.8}_{-1.5}$ &
$+0.1^{+0.3}_{-0.5}$
& $-1.0^{+0.6}_{-0.2}$ & $-1.4^{+1.9}_{-2.1}$ & $-6.3^{+2.4}_{-1.3}$ &
& $6.5^{+1.9}_{-2.5}$\\
\end{tabular}
\end{table}
\vbox{\begin{table}
\caption{Results for the lightest few negative-parity nucleon resonances of
each $J$ in the
N=3 band in the $\Delta\pi$ and $N\rho$ channels. Notation as in
Table~\ref{nNle2Npipi}.}\label{nNeq3Npipi}
\begin{tabular}{lrrrrrrr}
\multicolumn{1}{l} {model state}
& \multicolumn{1}{r}{$\Delta\pi$}
& \multicolumn{1}{r}{$\Delta\pi$}
& \multicolumn{1}{r}{$\sqrt{\Gamma^{\rm tot}_{\Delta\pi}}$}
& \multicolumn{1}{r}{$N\rho$}
& \multicolumn{1}{r}{$N\rho$}
& \multicolumn{1}{r}{$N\rho$}
& \multicolumn{1}{r}{$\sqrt{\Gamma^{\rm tot}_{N\rho}}$}\\
\multicolumn{1}{l} {$N\pi$ state/rating} & & & & & & &\\
\tableline
& $d$ & & & $s_{\textstyle{1\over2}}$ & $d_{\textstyle{3\over2}}$ & &\\
\cline{2-2}
\cline{5-6}
$[N{\textstyle{1\over2}}^-]_3(1945)$ & $-6.7^{+1.3}_{-1.5}$ & &
$6.7^{+1.5}_{-1.3}$
& $+2.3\pm 0.6$ & $-17.9^{+7.3}_{-3.8}$ & & $18.1^{+3.9}_{-7.3}$\\
$N{\textstyle{1\over2}}^-(2090)S_{11}$* & $-5.1\pm 5.9$ & $$ & $5.1\pm 5.9$ &
$+14.2\pm 4.3$
& $0.0\pm 2.0$ & $$ & $14.2\pm 4.3$\\

$[N{\textstyle{1\over2}}^-]_4(2030)$ & $-5.7^{+1.1}_{-1.3}$ & &
$5.7^{+1.3}_{-1.1}$
& $-0.5\pm 0.2$ & $-0.9\pm 0.3$ & & $1.1\pm 0.4$\\
$[N{\textstyle{1\over2}}^-]_5(2070)$ & $+13.1^{+3.3}_{-2.7}$ & &
$13.1^{+3.3}_{-2.7}$
& $+3.6^{+1.3}_{-1.2}$ & $-6.9^{+3.3}_{-1.3}$ & & $7.8^{+1.7}_{-3.5}$\\
$[N{\textstyle{1\over2}}^-]_6(2145)$ & $+1.0\pm 0.2$ & & $1.0\pm 0.2$ &
$-0.1^{+0.0}_{-0.1}$
& $+2.3^{+1.4}_{-0.6}$ & & $2.3^{+1.4}_{-0.6}$\\
\end{tabular}
\end{table}}
\vbox{\begin{table}
\caption{Results for nucleons in the $N=3$ band in the $\Delta(1600)\pi$, and
$N(1440)\pi$ channels. Notation as in Table~\ref{nNle2Npipi}.}
\label{nNeq3Npipi2}
\begin{tabular}{lrrrr}
\multicolumn{1}{l} {model state}
& \multicolumn{1}{r}{$\Delta(1600)\pi$}
& \multicolumn{1}{r}{$\Delta(1600)\pi$}
& \multicolumn{1}{r}{$\sqrt{\Gamma^{\rm tot}_{\Delta(1600)\pi}}$}
& \multicolumn{1}{r}{$N(1440)\pi$}\\
\multicolumn{1}{l} {$N\pi$ state/rating} & & & & \\
\tableline
& $d$ & & & \\
\cline{2-2}
$[N{\textstyle{1\over2}}^-]_3(1945)$ &$  6.4^{+3.4}_{-2.8}$  & &$
6.4^{+3.4}_{-2.8}$
& $  2.9\pm 0.4$ \\
$N{\textstyle{1\over2}}^-(2090)S_{11}$* & & & & $-11.1\pm 4.4$\\

$[N{\textstyle{1\over2}}^-]_4(2030)$ &$  4.2^{+2.5}_{-2.0}$  & &$
4.2^{+2.5}_{-2.0}$
&$  2.6^{+0.5}_{-0.3}$  \\
$[N{\textstyle{1\over2}}^-]_5(2070)$ &$ -1.2^{+0.3}_{-0.1}$  & &$
1.2^{+0.1}_{-0.3}$
&$ -5.4\pm 0.2$   \\
$[N{\textstyle{1\over2}}^-]_6(2145)$ &$ -0.1\pm 0.0$  & &$  0.1\pm 0.0$ &$
-0.3\pm 0.0$\\
\end{tabular}
\end{table}}
\begin{table}
\caption{Results in the $N\pi$ and $\Delta\pi$ channels for the lightest few
negative-parity $\Delta$ resonances
of each $J$ in the N=3 band, and
for the lightest few $\Delta$ resonances for $J^P$ values which
first appear in the N=4, 5 and 6 bands. Notation as in
Table~\ref{nNle2PS}.}\label{dNge3NDpi}
\begin{tabular}{lrrrr}
\multicolumn{1}{l} {model state}
& \multicolumn{1}{r}{$N\pi$}
& \multicolumn{1}{r}{$\Delta\pi$}
& \multicolumn{1}{r}{$\Delta\pi$}
& \multicolumn{1}{r}{$\sqrt{\Gamma^{\rm tot}_{\Delta\pi}}$}\\
\multicolumn{1}{l} {$N\pi$ state/rating} & & & &\\
\tableline
& & $d$ &\\
\cline{3-3}
$[\Delta{\textstyle{1\over2}}^-]_2(2035)$ & $1.2\pm 0.2$ & $+8.2^{+0.6}_{-0.7}$
&
& $8.2^{+0.6}_{-0.7}$\\
$\Delta{\textstyle{1\over2}}^-(1900)S_{31}$*** & $4.1\pm 2.2$ & $+6.4\pm 1.9$ &
$$
& $6.4\pm 1.9$\\

$[\Delta{\textstyle{1\over2}}^-]_3(2140)$ & $3.1^{+0.4}_{-1.1}$ &
$-4.4^{+0.8}_{-0.7}$ &
& $4.4^{+0.7}_{-0.8}$\\
$\Delta{\textstyle{1\over2}}^-(2150)S_{31}$* & $4.0\pm 1.5$\\

& & $s$ & $d$\\
\end{tabular}
\end{table}
%
\begin{table}
\caption{Results in the $N\rho$ channel for the lightest few negative-parity
$\Delta$ resonances
of each $J$ in the N=3 band, and for the lightest
few $\Delta$ resonances for $J^P$ values which first appear in the N=4, 5 and
6 bands. Notation as
in Table~\ref{nNle2Npipi}.}\label{dNge3Nrho}
\begin{tabular}{lrrrr}
\multicolumn{1}{l} {model state}
& \multicolumn{1}{r}{$N\rho$}
& \multicolumn{1}{r}{$N\rho$}
& \multicolumn{1}{r}{$N\rho$}
& \multicolumn{1}{r}{$\sqrt{\Gamma^{\rm tot}_{N\rho}}$}\\
\multicolumn{1}{l} {$N\pi$ state/rating} & & & &\\
\tableline
& $s_{\textstyle{1\over2}}$ & $d_{\textstyle{3\over2}}$ & &\\
\cline{2-3}
$[\Delta{\textstyle{1\over2}}^-]_2(2035)$ & $+2.5\pm 0.6$ &
$+1.5^{+0.5}_{-0.3}$ &
& $2.9^{+0.8}_{-0.6}$\\
$\Delta{\textstyle{1\over2}}^-(1900)S_{31}$*** & $-3.5\pm 2.7$ & $-9.3\pm 1.7$
& $$
& $9.9\pm 1.9$\\

$[\Delta{\textstyle{1\over2}}^-]_3(2140)$ & $-2.2\pm 0.6$ &
$+2.3^{+1.0}_{-0.4}$ &
& $3.2^{+1.2}_{-0.7}$\\
$\Delta{\textstyle{1\over2}}^-(2150)S_{31}$*\\

& $d_{\textstyle{1\over2}}$ & $s_{\textstyle{3\over2}}$ &
$d_{\textstyle{3\over2}}$ &\\
\end{tabular}
\end{table}
\begin{table}
\caption{Results in the $\Delta(1600)\pi$, and $N(1440)\pi$ channels for the
lightest few
negative-parity $\Delta$ resonances in the $N=3$ band, and
for the lightest few $\Delta$ resonances for $J^P$ values that first appear in
the $N=4, 5$ and 6 bands. Notation as in Table~\ref{nNle2Npipi}.}
\label{dNge2Npipi2}
\begin{tabular}{lrrrr}
\multicolumn{1}{l} {model state}
& \multicolumn{1}{r}{$\Delta(1600)\pi$}
& \multicolumn{1}{r}{$\Delta(1600)\pi$}
& \multicolumn{1}{r}{$\sqrt{\Gamma^{\rm tot}_{\Delta(1600)\pi}}$}
& \multicolumn{1}{r}{$N(1440)\pi$}\\
\multicolumn{1}{l} {$N\pi$ state/rating} & & & & \\
\tableline
& $d$ & & & \\
\cline{2-2}
$[\Delta{\textstyle{1\over2}}^-]_2(2035)$ & $ -4.0^{+0.9}_{-1.0}$ &  &$
4.0^{+1.0}_{-0.9}$
&$  1.8\pm 0.1$\\
$\Delta{\textstyle{1\over2}}^-(1900)S_{31}$*** & & & & $-4.1\pm 2.8$\\

$[\Delta{\textstyle{1\over2}}^-]_3(2140)$ & $ -1.4^{+0.8}_{-1.3}$ &  &$
1.4^{+1.3}_{-0.8}$
&$ -5.1^{+0.7}_{-0.1}$\\
$\Delta{\textstyle{1\over2}}^-(2150)S_{31}$* & \\

& $s$ & $d$ &\\
\end{tabular}
\end{table}
\begin{table}
\caption{Results in the $N\pi$, $N\eta$, and $N\omega$ channels for the
lightest few
nucleon resonances for $J^P$ values which first appear in the
N=4, 5 and 6 bands. Notation as in Table~\ref{nNle2PS}.}\label{nNge4PS}
\begin{tabular}{lrrrrrrr}
\multicolumn{1}{l} {model state}
& \multicolumn{1}{r}{$N\pi$}
& \multicolumn{1}{r}{$N\eta$}
& \multicolumn{1}{r}{$N\eta^\prime$}
& \multicolumn{1}{r}{$N\omega$}
& \multicolumn{1}{r}{$N\omega$}
& \multicolumn{1}{r}{$N\omega$}
& \multicolumn{1}{r}{$\sqrt{\Gamma^{\rm tot}_{N\omega}}$}\\
\multicolumn{1}{l} {$N\pi$ state/rating}\\
\tableline
& & & & $p_{\textstyle{1\over2}}$ & $p_{\textstyle{3\over2}}$ &\\
\cline{5-6}
$[N{\textstyle{1\over2}}^+]_6(2065)$ & $7.7^{+2.4}_{-2.9}$ & $+1.1\pm 1.1$ &
$-0.6\pm 0.5$
& $+0.6^{+0.1}_{-0.2}$ & $-0.4^{+2.9}_{-3.7}$ & & $0.7^{+3.5}_{-0.7}$\\
$N{\textstyle{1\over2}}^+(2100)P_{11}$* & $5.0\pm 2.0$\\

$[N{\textstyle{1\over2}}^+]_7(2210)$ & $0.3^{+0.2}_{-0.1}$ &
$-1.1^{+0.8}_{-0.7}$
& $+0.6^{+0.5}_{-0.9}$ & $+2.1\pm 1.8$ & $-1.5\pm 0.3$ &
& $2.5^{+1.7}_{-1.3}$\\

& & & & $f_{\textstyle{1\over2}}$ & $f_{\textstyle{3\over2}}$ &
$h_{\textstyle{3\over2}}$\\
\cline{5-7}
$[N{\textstyle{7\over2}}^+]_2(2390)$ & $4.9^{+0.1}_{-0.4}$ & $+1.7\pm 0.1$
& $+1.3^{+0.3}_{-0.4}$ & $-0.8\pm 0.1$ & $+2.1^{+0.2}_{-0.4}$
& $+2.0^{+1.2}_{-0.9}$ & $3.0^{+1.0}_{-0.9}$\\
\end{tabular}
\end{table}
\begin{table}
\caption{Results in the $\Delta\pi$ and $N\rho$ channels for the lightest few
nucleon resonances
for $J^P$ values which first appear in the N=4, 5
and 6 bands. Notation as in Table~\ref{nNle2Npipi}.}\label{nNge4Npipi}
\begin{tabular}{lrrrrrrr}
\multicolumn{1}{l} {model state}
& \multicolumn{1}{r}{$\Delta\pi$}
& \multicolumn{1}{r}{$\Delta\pi$}
& \multicolumn{1}{r}{$\sqrt{\Gamma^{\rm tot}_{\Delta\pi}}$}
& \multicolumn{1}{r}{$N\rho$}
& \multicolumn{1}{r}{$N\rho$}
& \multicolumn{1}{r}{$N\rho$}
& \multicolumn{1}{r}{$\sqrt{\Gamma^{\rm tot}_{N\rho}}$}\\
\multicolumn{1}{l} {$N\pi$ state/rating} & & & & & & &\\
\tableline
& $p$ & & & $p_{\textstyle{1\over2}}$ & $p_{\textstyle{3\over2}}$ &\\
\cline{2-2}
\cline{5-6}
$[N{\textstyle{1\over2}}^+]_6(2065)$ & $-1.9^{+1.0}_{-2.8}$ & &
$1.9^{+2.8}_{-1.0}$
& $+1.4^{+3.1}_{-0.1}$ & $+5.7\pm 2.9$ & & $5.9^{+3.9}_{-2.8}$\\
$N{\textstyle{1\over2}}^+(2100)P_{11}$*\\

$[N{\textstyle{1\over2}}^+]_7(2210)$ & $-7.0^{+4.3}_{-6.4}$ & &
$7.0^{+6.4}_{-4.3}$
& $-2.0^{+0.7}_{-1.8}$ & $-2.6\pm 1.2$ & & $3.3^{+2.1}_{-1.3}$\\

& $f$ & $h$ & & $f_{\textstyle{1\over2}}$ & $f_{\textstyle{3\over2}}$ &
$h_{\textstyle{3\over2}}$\\
\cline{2-3}
\cline{5-7}
$[N{\textstyle{7\over2}}^+]_2(2390)$ & $+0.1\pm 0.0$ & $-1.8^{+0.9}_{-1.2}$
& $1.8^{+1.2}_{-0.9}$ & $-5.2^{+2.1}_{-0.0}$ & $+4.7^{+0.0}_{-1.9}$
& $+6.4^{+2.5}_{-3.7}$ & $9.5^{+1.8}_{-4.6}$\\
\end{tabular}
\end{table}
\vbox{\begin{table}
\caption{Results in the $\Delta(1600)\pi$ and $N(1440)\pi$ channels for the
lightest few nucleon resonances for $J^P$ values which first appear in the N=4,
5 and 6 bands. Notation as in Table~\ref{nNle2Npipi}.}
\label{nNge4Npipi2}
\begin{tabular}{lrrrr}
\multicolumn{1}{l} {model state}
& \multicolumn{1}{r}{$\Delta(1600)\pi$}
& \multicolumn{1}{r}{$\Delta(1600)\pi$}
& \multicolumn{1}{r}{$\sqrt{\Gamma^{\rm tot}_{\Delta(1600)\pi}}$}
& \multicolumn{1}{r}{$N(1440)\pi$}\\
\multicolumn{1}{l} {$N\pi$ state/rating} & & & & \\
\tableline
& $p$ & & & \\
\cline{2-2}

$[N{\textstyle{1\over2}}^+]_6(2065)$ &$  9.1^{+1.5}_{-2.5}$  & &$
9.1^{+1.5}_{-2.5}$
&$  7.9\pm 0.4$   \\
$N{\textstyle{1\over2}}^+(2100)P_{11}$*\\

$[N{\textstyle{1\over2}}^+]_7(2210)$ & $  8.1\pm 0.5$  & &$  8.1\pm 0.5$
&$  2.1^{+0.0}_{-0.3}$\\

& $f$ & $h$ & & \\
\cline{2-3}

$[N{\textstyle{7\over2}}^+]_2(2390)$ & $ -0.1\pm 0.0$   &$  0.5^{+0.4}_{-0.3}$
&$  0.6^{+0.4}_{-0.3}$&$ -1.6\pm 0.1$ \\
\end{tabular}
\end{table}}


\begin{references}
\bibitem{miss-undisc}{We adopt the conventional distinction between `missing'
and
`undiscovered' baryon states: the former refers to states predicted by
nonrelativistic quark models
to lie in the first band of positive-parity excited states (states which
are predominantly $N=2$ band when the wavefunction is expanded in a harmonic
oscillator basis); the
latter refers to all other states predicted by the model but unseen (in the
$N=3$ bands and beyond).}
\end{references}
\end{document}